\documentclass[iop]{emulateapj}

\usepackage{graphicx} 
\usepackage{amsmath}
\usepackage{appendix}
\usepackage{amssymb}
\usepackage{aasmacros}
\usepackage{natbib}
\usepackage{latexsym}
\usepackage{threeparttable} 
\usepackage{subfigure} 
\usepackage[figoff]{figcaps}
\usepackage{ccaption}
\usepackage{hyperref}
\captiondelim{ }

\begin{document}

\pagestyle{plain}
\title{Binary Asteroid Encounters with Terrestrial Planets: Timescales and Effects}
\author{Julia Fang\altaffilmark{1} and Jean-Luc Margot\altaffilmark{1,2}}

\altaffiltext{1}{Department of Physics and Astronomy, University of California, Los Angeles, CA 90095, USA}
\altaffiltext{2}{Department of Earth and Space Sciences, University of California, Los Angeles, CA 90095, USA}

\begin{abstract}

Many asteroids that make close encounters with terrestrial planets
are in a binary configuration.  Here we calculate the relevant
encounter timescales and investigate the effects of encounters on a
binary's mutual orbit. We use a combination of analytical and
numerical approaches with a wide range of initial conditions.  Our
test cases include generic binaries with close, moderate, and wide
separations, as well as seven well-characterized near-Earth binaries.
We find that close approaches ($<$10 Earth radii) occur for almost
all binaries on 1-10 million year timescales.  At such distances, our
results suggest substantial modifications to a binary's semi-major
axis, eccentricity, and inclination, which we quantify.  Encounters
within 30 Earth radii typically occur on sub-million year timescales and
significantly affect the wider binaries.  Important processes in the
lives of near-Earth binaries, such as tidal and radiative evolution,
can be altered or stopped by planetary encounters.

\end{abstract}
\maketitle

\section{Introduction} \label{intro}

Near-Earth asteroids (NEAs) live in a dynamic environment that
includes gravitational encounters with planets and the Sun as well as other
non-gravitational perturbations, with average dynamical lifetimes on
the order of a few million years \citep{bott02}.  They are replenished
from source regions in the main belt of asteroids, which include
strong resonances with Jupiter and Saturn. Radar and light curve
studies by \citet{marg02} and \citet{prav06} have determined that
approximately 15\% of these short-lived objects larger than 200~m in
diameter are in a binary configuration, with a primary and secondary
orbiting their common center of mass.
The mutual orbits of these binary asteroids are affected by tides, by
external perturbations such as the radiative binary YORP effect
\citep{cuk05,cuk07}, and by close encounters with terrestrial
planets. In this work, we investigate the effect of close planetary
encounters on the mutual binary orbits of NEA systems by studying
changes in semi-major axis, eccentricity, and inclination. We also
calculate the frequency of such terrestrial planet flybys.

Early work on the subject of planetary encounters with NEA binaries
was accomplished by \citet{fari92b} and \citet{fari93}. 
They employed analytical estimates and Monte Carlo techniques to
estimate the change in orbital energy and angular momentum due to an
encounter. Recent work by \citet{fang11} examined the effect of close
planetary encounters on the mutual orbits of NEA triples 2001 SN263
and 1994 CC. 
They found that scattering events by terrestrial planets can excite
the eccentricities as well as mutual inclinations of the satellites'
orbits to currently observed values on million-year timescales. During
close approaches with Earth, the outer satellites in both triple
systems can have their orbital eccentricities excited to values of at
least 0.2 as far away as encounter distances of $\sim$40 Earth radii
for 2001 SN263 and $\sim$50 Earth radii for 1994 CC. Since the orbital
effects of encounters on triples have been previously examined, this
present study focuses only on binary systems. Other recent studies
have invoked planetary encounters to explain the rotational dynamics
of asteroids \citep{sche04,shar06}, the tidal disruption of rubble
pile asteroids to form binaries \citep{wals06}, and the re-surfacing
of NEAs \citep{nesv10,binz10}. Prior studies include \citet{bott96b,
  bott96a}, \citet{asph96}, and \citet{rich98}.

In this work, we perform N-body simulations and employ analytical
expressions to evaluate a binary's orbital changes due to a close
planetary encounter with Earth. Initial conditions include 3 generic
cases (a ``close binary'' with a separation of 4 primary radii, a
``moderately-separated binary'' with a separation of 8 primary radii,
and a ``wide binary'' with a separation of 16 primary radii) as well
as parameters drawn from known, well-characterized NEA binaries. By
``well-characterized'' we mean a binary for which the system mass,
semi-major axis, eccentricity, and approximate component sizes are known - in
practice this corresponds to a subset of the radar-observed binaries.
Table \ref{nea} shows a compilation of these binaries and their
parameters such as primary size, primary mass, and primary-secondary
separation. In Section \ref{nea_single}, we perform single flyby
simulations and use analytical equations to determine the orbital
effects of a close planetary encounter, by thoroughly examining a
variety of encounter geometries, distances, and velocities. We study
the changes in the semi-major axis, eccentricity, and inclination of
the binary orbit. In Section \ref{longterm}, we present long-term
simulations with test particles and planets to calculate encounter
timescales for observed NEA binaries, whose encounter frequencies are
strongly dependent on their individual evolutionary histories. We
discuss and summarize this study in Section \ref{conclusion}.

\def\arraystretch{1.4}
\begin{deluxetable*}{lrrrrrrr}
\tablecolumns{8}
\tablecaption{Well-Characterized Near-Earth Binaries \label{nea}}
\startdata
\hline \hline
System		& $R_p$ (km) 	& $M_p$ (kg) & $a$ (km) & $a/R_p$ & $a_{\odot}$ (AU) & $e_{\odot}$ & $i_{\odot}$ (deg) \\
\hline
(185851) 2000 DP107\footnotemark[1]	& 0.40 & 4.38 $\times$ $10^{11}$ & 2.62 & 6.6 & 1.37 & 0.38 & 8.67	\\
(66391) 1999 KW4\footnotemark[2] 	& 0.66 & 2.35 $\times$ $10^{12}$ & 2.55 & 3.9 & 0.64 & 0.69 & 38.89 \\
(276049) 2002 CE26\footnotemark[3] 	& 1.75 & 2.17 $\times$ $10^{13}$ & 4.87 & 2.8 & 2.23 & 0.56 & 47.31	\\
2004 DC\footnotemark[4] 			& 0.17 & 3.57 $\times$ $10^{10}$ & 0.75 & 4.4 & 1.63 & 0.40 & 19.45	\\
(164121) 2003 YT1\footnotemark[5] 	& 0.55 & 1.89 $\times$ $10^{12}$ & 3.93 & 7.1 & 1.11 & 0.29 & 44.06 \\
(65803) Didymos\footnotemark[6] 	& 0.40 & 5.24 $\times$ $10^{11}$ & 1.18 & 3.0 & 1.64 & 0.38 & 3.41	\\
(35107) 1991 VH\footnotemark[7] 	& 0.60 & 1.40 $\times$ $10^{12}$ & 3.26 & 5.4 & 1.14 & 0.14 & 13.91
\enddata

\tablenotetext{}{Well-characterized near-Earth binaries and their
  parameters are compiled in this table, including the radius of the
  primary $R_p$, mass of the primary $M_p$, semi-major axis $a$
  (in units of km), and semi-major axis divided by the
  primary radius $a/R_p$. Rough uncertainties in binary parameters are 
  $\sim$20\% in sizes, $\sim$10\% in masses, and $\lesssim$5$-$10\% in 
  semi-major axes. Heliocentric orbital data are given for the semi-major
  axis $a_{\odot}$, eccentricity $e_{\odot}$, and inclination $i_{\odot}$
  with respect to the ecliptic.}

\footnotetext[1]{\citet{marg02}} 
\footnotetext[2]{\citet{ostr06}}
\footnotetext[3]{\citet{shep06}}
\footnotetext[4]{\citet{tayl08b}}
\footnotetext[5]{\citet{nola04}}
\footnotetext[6]{\citet{benn10}}
\footnotetext[7]{\citet{marg08,prav06}}
\end{deluxetable*}


\section{Effect of a Single Planetary Encounter} \label{nea_single}

\subsection{Methods}

We perform N-body numerical integrations to investigate the effects of
a close approach to an Earth-mass planet on a binary's mutual
orbit. We use a Bulirsch-Stoer algorithm from an N-body numerical
integration package, {\em Mercury} \citep{cham99}. Our simulations
incorporate 3 massive bodies (all assumed to be spherical), which include an
asteroid binary and an Earth-mass perturber on a hyperbolic trajectory
at various close encounter distances. 
We explore a wide range of initial conditions in encounter distance
and velocity, which we describe in detail below.  For each pair of
encounter distance and velocity, systematic simulations are performed
with nearly 7,000 permutations of the following mutual orbital
parameters: inclination, longitude of the ascending node, and mean
anomaly.  Our test cases include all of the NEA binaries shown in Table
\ref{nea} as well as a few generic cases with a typical rubble pile
density of 2 g cm$^{-3}$ and a primary radius of 0.5 km: a ``close
binary'' with a separation of 4 primary radii, a
``moderately-separated binary'' with a separation of 8 primary radii,
and a ``wide binary'' with a separation of 16 primary radii.

\begin{figure}[htb]
	\centering
	\includegraphics[width=3.3in]{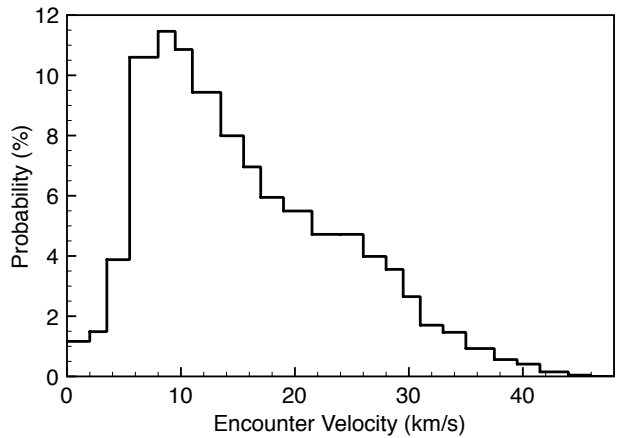}
	\caption{This probability distribution of expected encounter
          velocities $v_{\infty}$ of NEAs with Earth is obtained from
          numerical simulations of asteroid migration from source
          regions (Mars-crossing region, $\nu_6$ resonance, and 3:1
          mean-motion resonance with Jupiter) in or adjacent to the
          main belt (Bill Bottke and Kevin Walsh, personal
          communication, 2011). \label{nea_vinf}}
\end{figure}

The hyperbolic encounter velocity $v_{\infty}$, defined as the
relative speed between the binary and Earth at infinity, can be
described by a probability distribution of relative velocities as
shown in Figure \ref{nea_vinf}. As a result, our simulations
appropriately cover a range of $v_{\infty}$ from 8$-$24 km~s$^{-1}$
with increments of 4 km~s$^{-1}$. We also systematically repeat the
ensemble of simulations for various close encounter distances whose
range is dependent on the binary's semi-major axis, and with typical
increments of 1$-$2 $R_{\Earth}$.

In this paper, the ``encounter distance'' $q$ is defined as the
distance of closest approach and the ``impact parameter'' $b$ is the
hypothetical encounter distance that would result in the absence of
gravitational focusing. The impact parameter can be described as the
perpendicular distance between the non-focused path of the perturber 
and the binary.
The usual definition relating impact parameter $b$ and encounter
distance $q$ is $b^2 = q^2 (1 + (2 G M_{\Earth})/(q v_{\infty}^2))$,
where $G$ is the gravitational constant, $M_{\Earth}$ is the mass of
Earth, and $v_{\infty}$ is the encounter velocity at infinity. For
most cases we consider here, the difference between $b$ and $q$ is
less than 1 Earth radius.

All of the simulations begin with asteroid binaries in circular
orbits. After a scattering encounter, possible outcomes include (a) an
intact, stable binary system, or instability marked by (b) collision
between any two bodies or (c) ejection of the secondary from the
binary system. Encounter results are recorded for changes in
semi-major axis, eccentricity, and inclination in stable encounters
only. Stable systems are defined as binaries with no collisions nor
ejections of the secondary.

The output from numerical simulations is compared with approximate
analytical results. We consider close planetary encounters with NEA
binaries to be impulsive events, defined as swift encounters where the
planet's interaction with the binary is much shorter than the orbital
period of the binary.
The encounter delivers an impulse representing a shift in velocity of
the binary's orbit, and this impulse approximation is valid in the
domain where $q/v_{\infty} \ll P$,
where $P$ is the binary's mutual orbital period. 
For such impulsive encounters, the change in a
binary's orbital elements has been analytically approximated by
previous studies. \citet{hegg96} derived the change in a binary's
eccentricity, \citet{coll08} derived the change in both eccentricity
and inclination, and \citet{fari93} and \citet{chau95} derived the
change in energy (which we can relate to a change in semi-major
axis). The typical change in a binary's orbital elements after
averaging over all encounter geometries (which include stable and
unstable encounters) can be estimated as
\begin{equation} \label{delta_a}
	\Delta a \approx 1.48 \sqrt{\dfrac{G}{M}} \dfrac{ M_{\Earth} a^{5/2} }{v_{\infty} q^2}
\end{equation}
\begin{equation} \label{delta_e}
	\Delta e \approx 1.89 \sqrt{\dfrac{G}{M}} \dfrac{ M_{\Earth} a^{3/2} }{v_{\infty} q^2}
\end{equation}
\begin{equation} \label{delta_i}
	\Delta i \approx 0.75 \sqrt{\dfrac{G}{M}} \dfrac{ M_{\Earth} a^{3/2} }{v_{\infty} q^2}
\end{equation}
where $G$ is the gravitational constant, $M_{\Earth}$ is the mass of
Earth, $v_{\infty}$ is the encounter velocity, and $q$ is the
encounter distance. The binary's semi-major axis is represented by
$a$, its eccentricity is $e$, its inclination is $i$, and its system
mass is $M$. The constants in front of Equations \ref{delta_e} and
\ref{delta_i} represent the averaging of encounter geometries, and
have been calculated by \citet{coll08}. In their paper, we note there is a factor
of 2 missing in the denominator of Equation A3.
By analogy, we obtain the constant in front of Equation \ref{delta_a}
by scaling the analytical curve to our numerical simulation results.

\subsection{Results}

\begin{figure}[htb]
	\centering \includegraphics[scale=0.35]{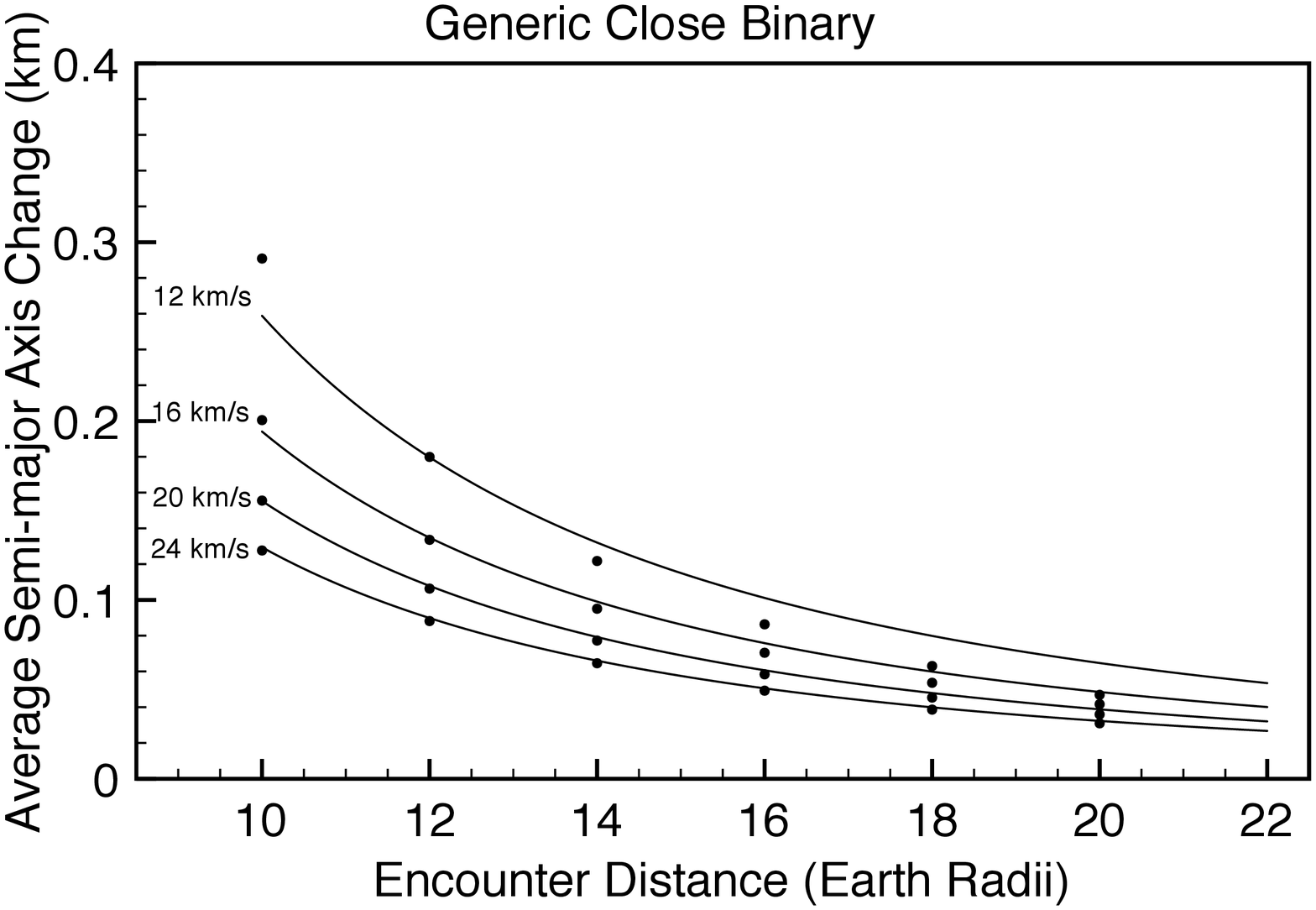}
	\includegraphics[scale=0.35]{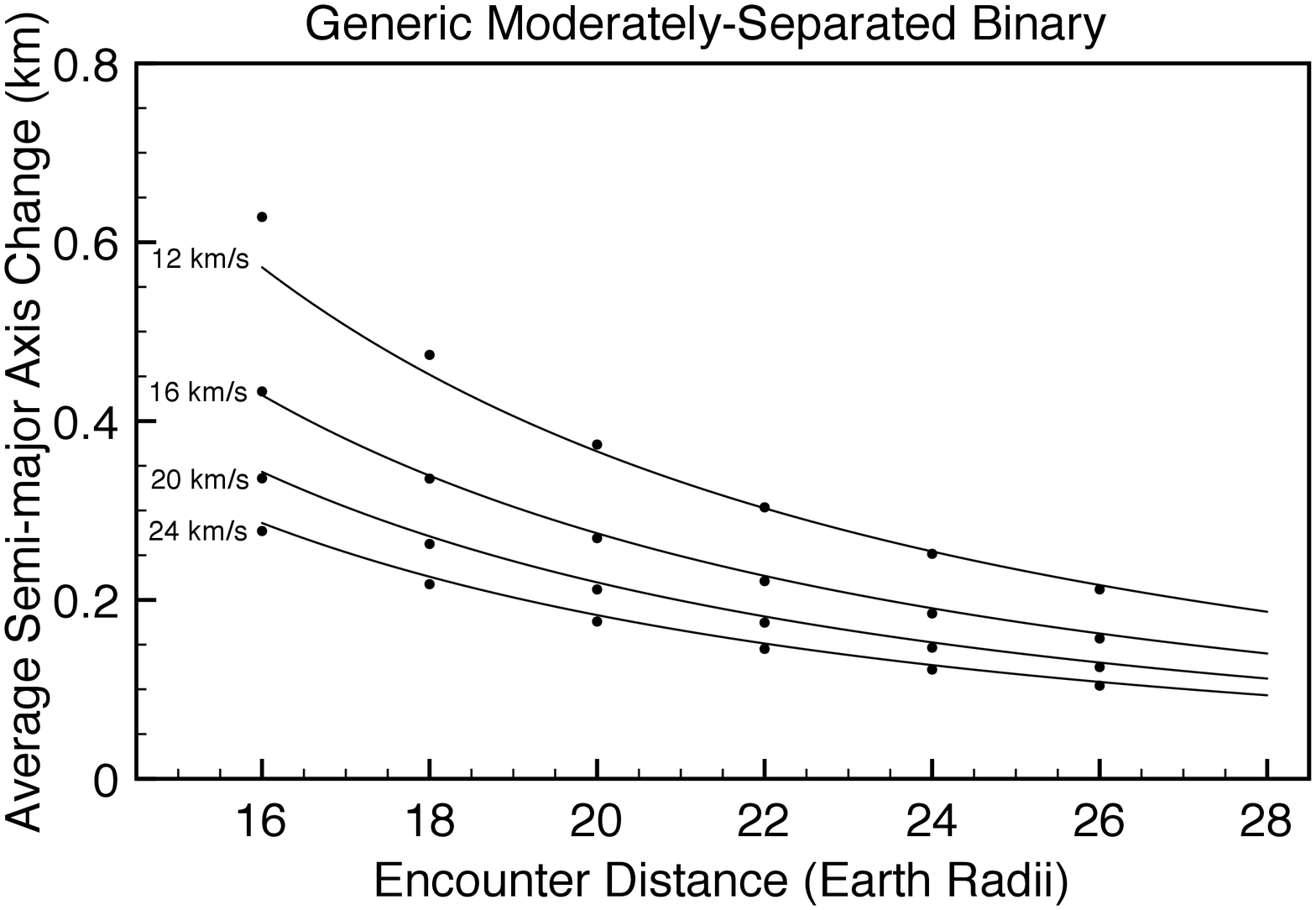}
	\includegraphics[scale=0.35]{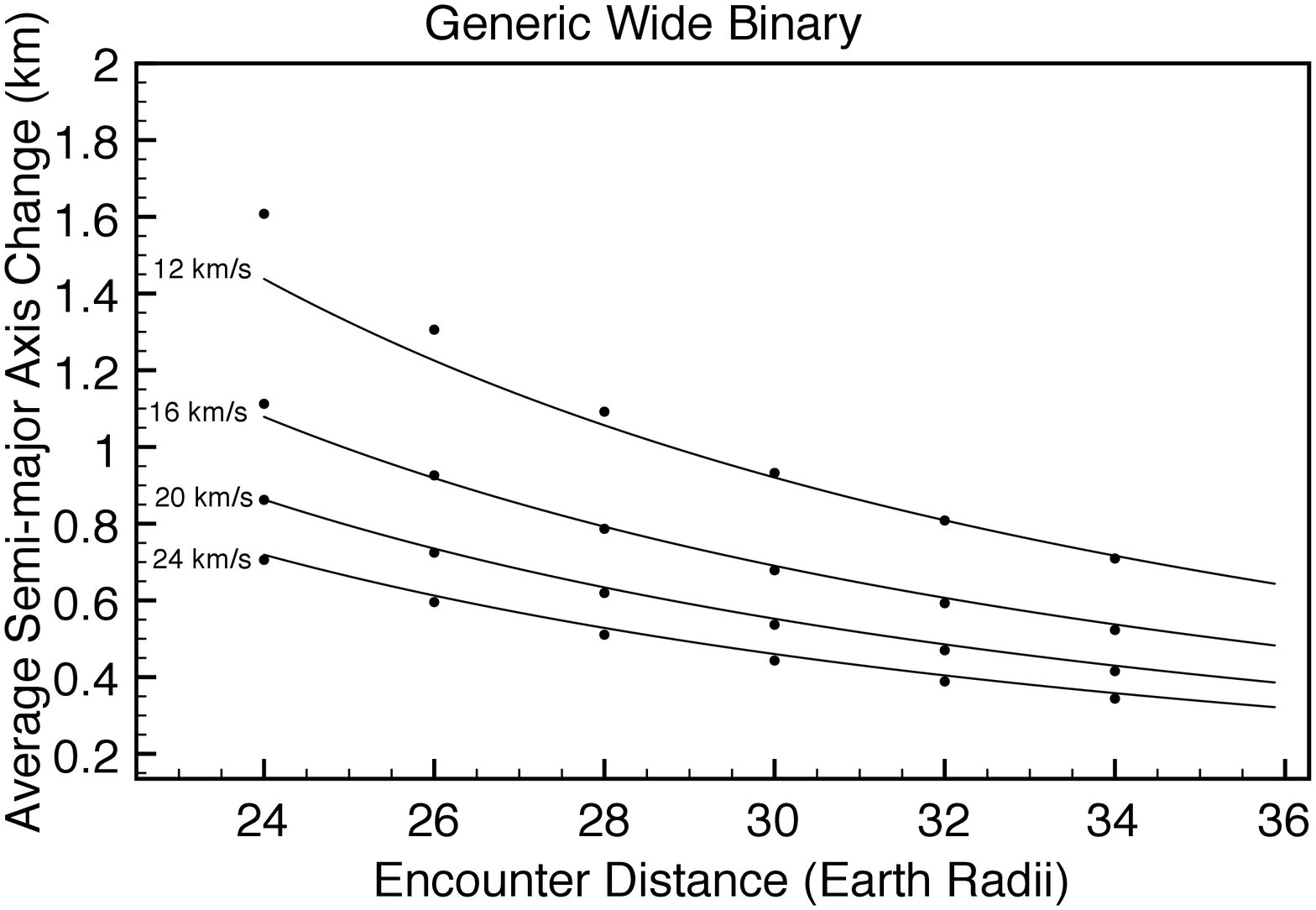}
	\caption{The change in a binary's semi-major axis is shown as
	a function of encounter distance and velocity for 3 types of
	binaries: close binaries, moderately-separated binaries, and
	wide binaries. Results from numerical simulations are shown
	as dots and analytical calculations are depicted by
	solid lines. \label{delta_a_plots}}
\end{figure}

The change in semi-major axis due to a planetary flyby is shown for
three generic cases presented in Figure \ref{delta_a_plots}. Output
from numerical simulations are shown as points and represent the mean
values of the change (after taking the absolute value) in semi-major
axis resulting from stable encounters. The curves represent analytical
estimates, which provide a reasonable match to our numerical
results. We show the averages changes in semi-major axis for
$v_{\infty}$ = 12, 16, 20, and 24 km~s$^{-1}$. The results using a
$v_{\infty}$ of 8 km~s$^{-1}$ are not shown because the resulting
spread in the post-encounter semi-major axis is very broad (e.g.~at an
encounter distance of 10 Earth radii, the standard deviation is
$\sim$2.1 km) and not accurately portrayed by a single value. 
We wish to show orbital effects of only stable encounters;
the range of close encounter distances for each binary type shown
in this figure is determined by the distances
at which all encounters resulted in stable binaries. Since the
3 types of binaries have different separations, they will
have different ranges of encounter distances at which stable
encounters occur.
These plots show that larger binary separations produce
greater changes in semi-major axis at any given encounter distance
because wider binaries are less tightly bound and thus more susceptible
to passing perturbers.

\begin{figure}[htb]
	\centering
	\includegraphics[scale=0.35]{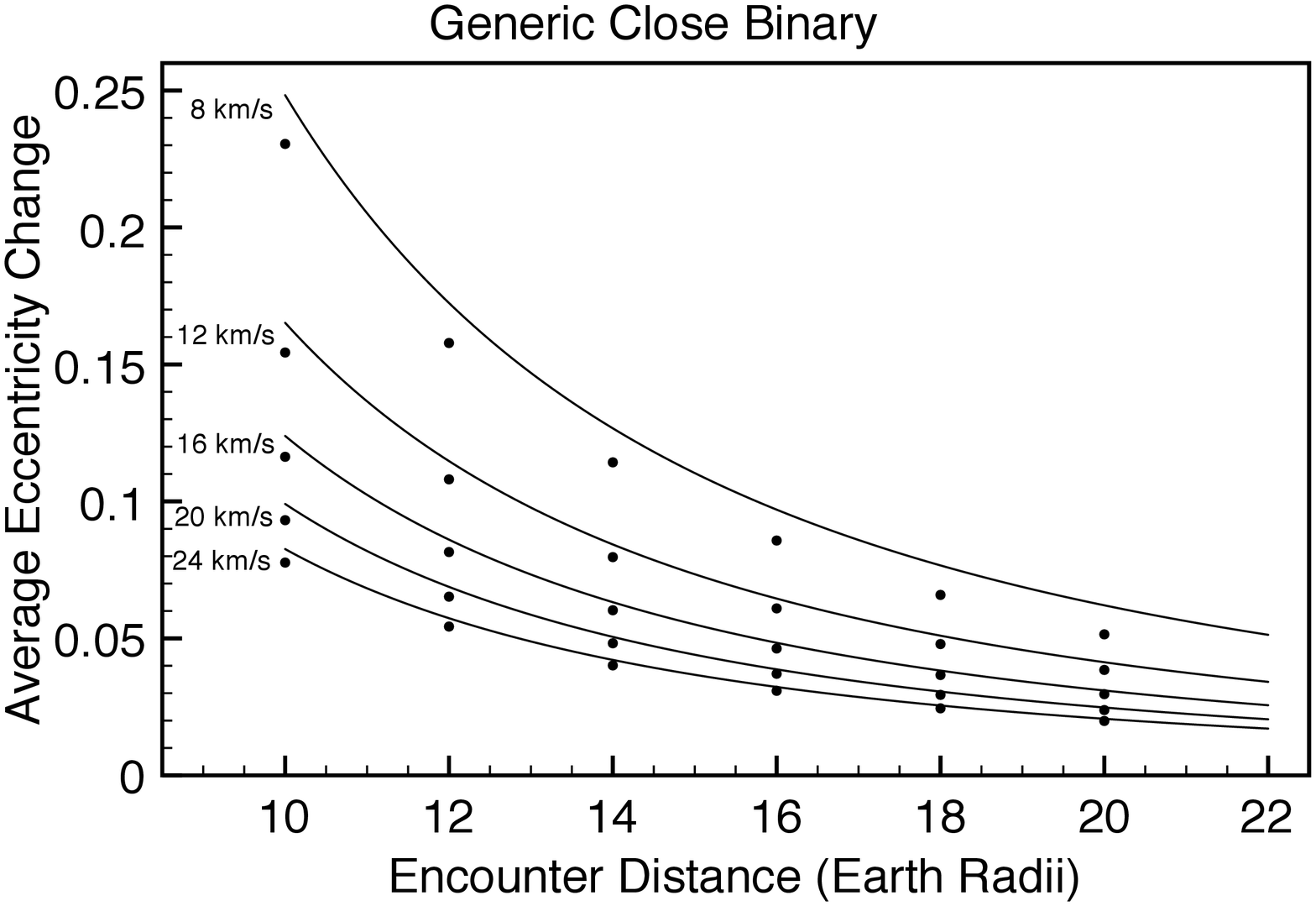}
	\includegraphics[scale=0.35]{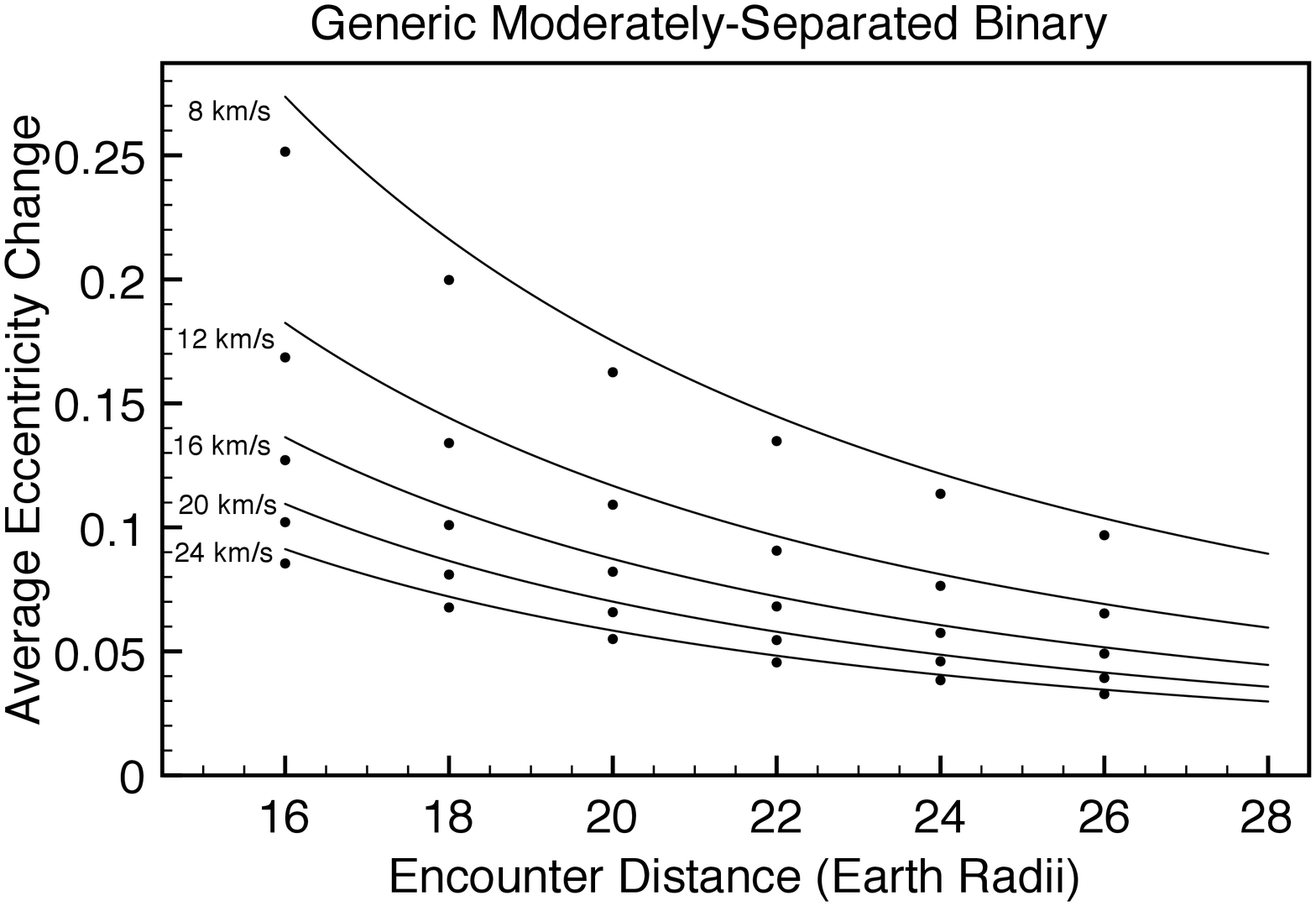}
	\includegraphics[scale=0.35]{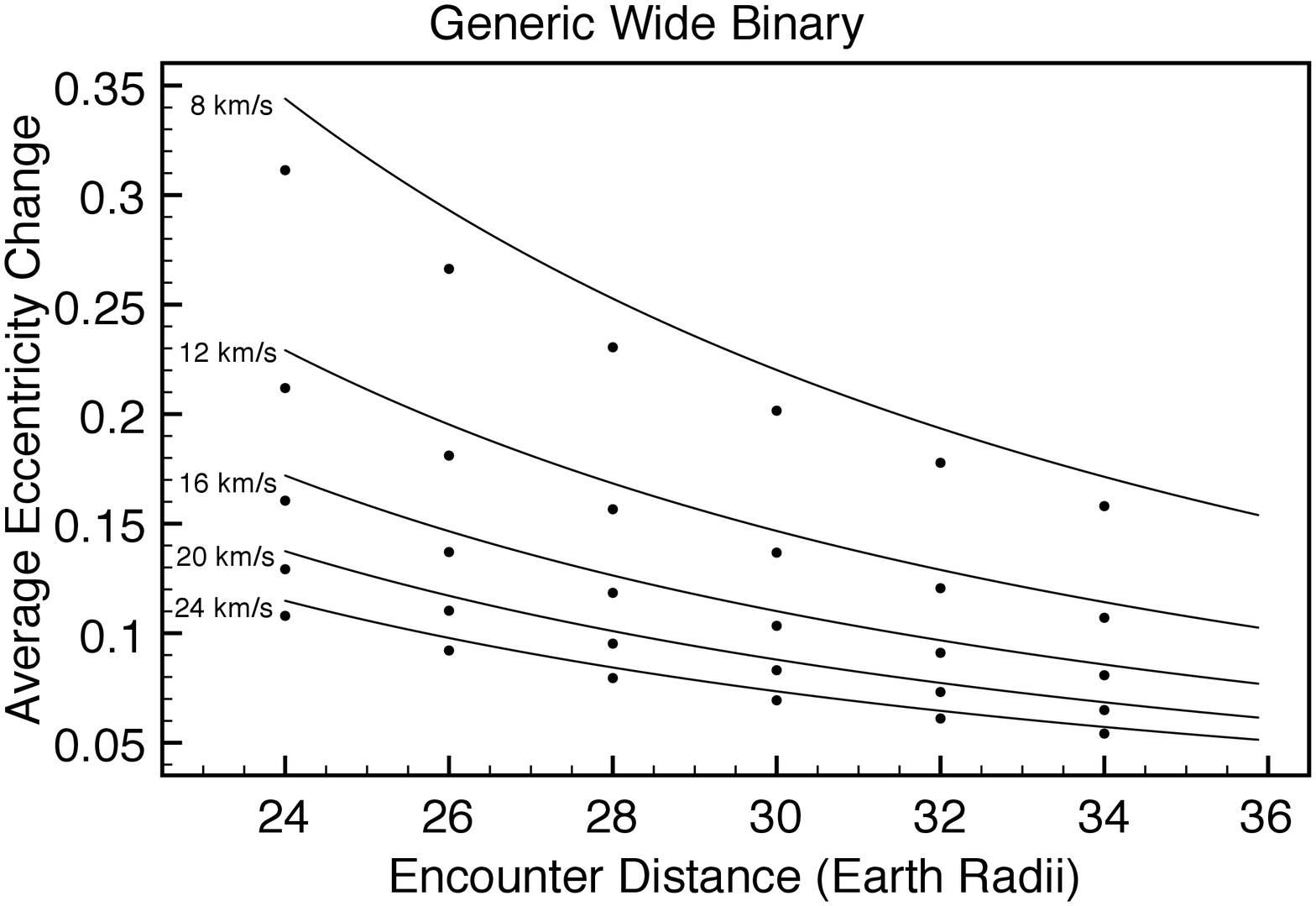}
	\caption{The change in a binary's eccentricity is shown as a
	function of encounter distance and velocity for 3 types of
	binaries: close binaries, moderately-separated binaries, and
	wide binaries. Results from numerical simulations are shown
	as dots and analytical calculations are depicted by
	solid lines. \label{delta_e_plots}}
\end{figure}

\begin{figure}[htb]
	\centering
	\includegraphics[scale=0.35]{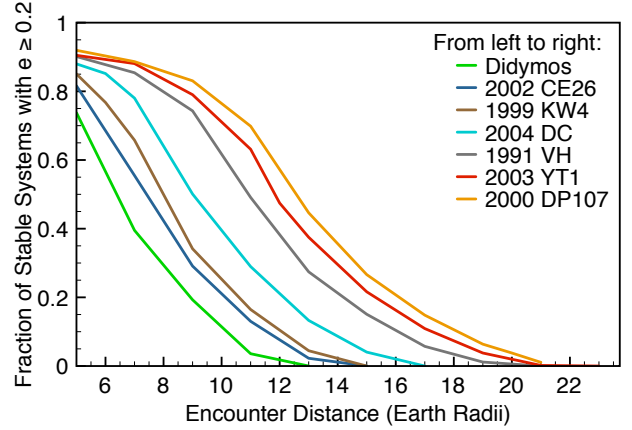}
	\caption{Using orbital and physical parameters from the sample
	of NEA binaries given in Table \ref{nea}, we show the fraction
	of stable systems (no ejections nor collisions) with excited
	eccentricities ($e$ $\geq$ 0.2) as a function of encounter
	distance for $v_{\infty}$ = 12 km~s$^{-1}$. \label{neaexcite}}
\end{figure}

\begin{figure}[thb]
	\centering
	\includegraphics[width=3.3in]{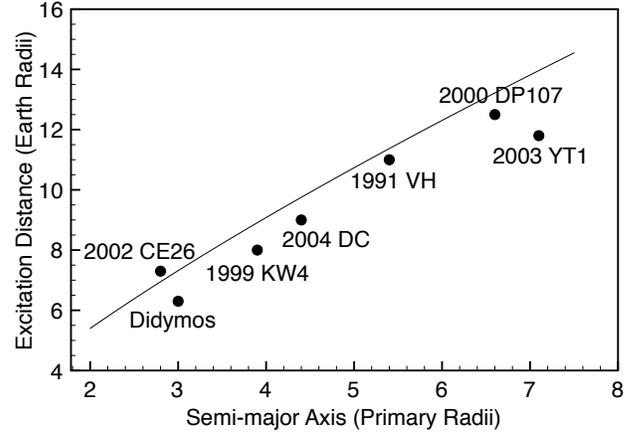}
	\caption{The eccentricity excitation distance (defined as the
	close encounter distance where 50\% of stable NEA binary
	systems showed $e$ $\geq$ 0.2) is given as a function of the
	secondary's semi-major axis in units of primary radii. The
	curve shows the result obtained from the analytical
	expression.  This case is for $v_{\infty}$ of 12 km~s$^{-1}$.
	\label{a_excite}}
\end{figure}

\begin{figure}[htb]
	\centering
	\includegraphics[scale=0.35]{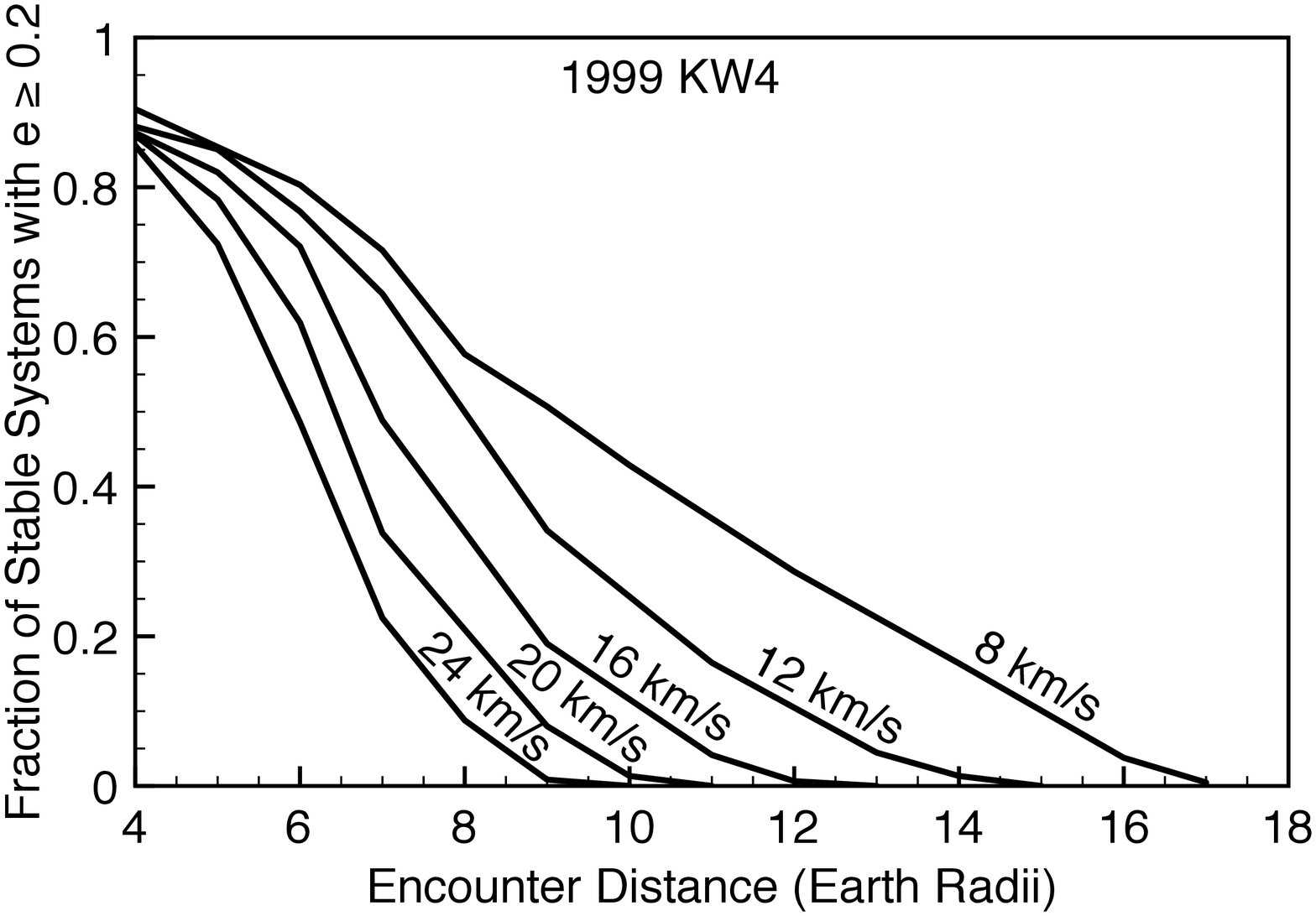}
	\includegraphics[scale=0.35]{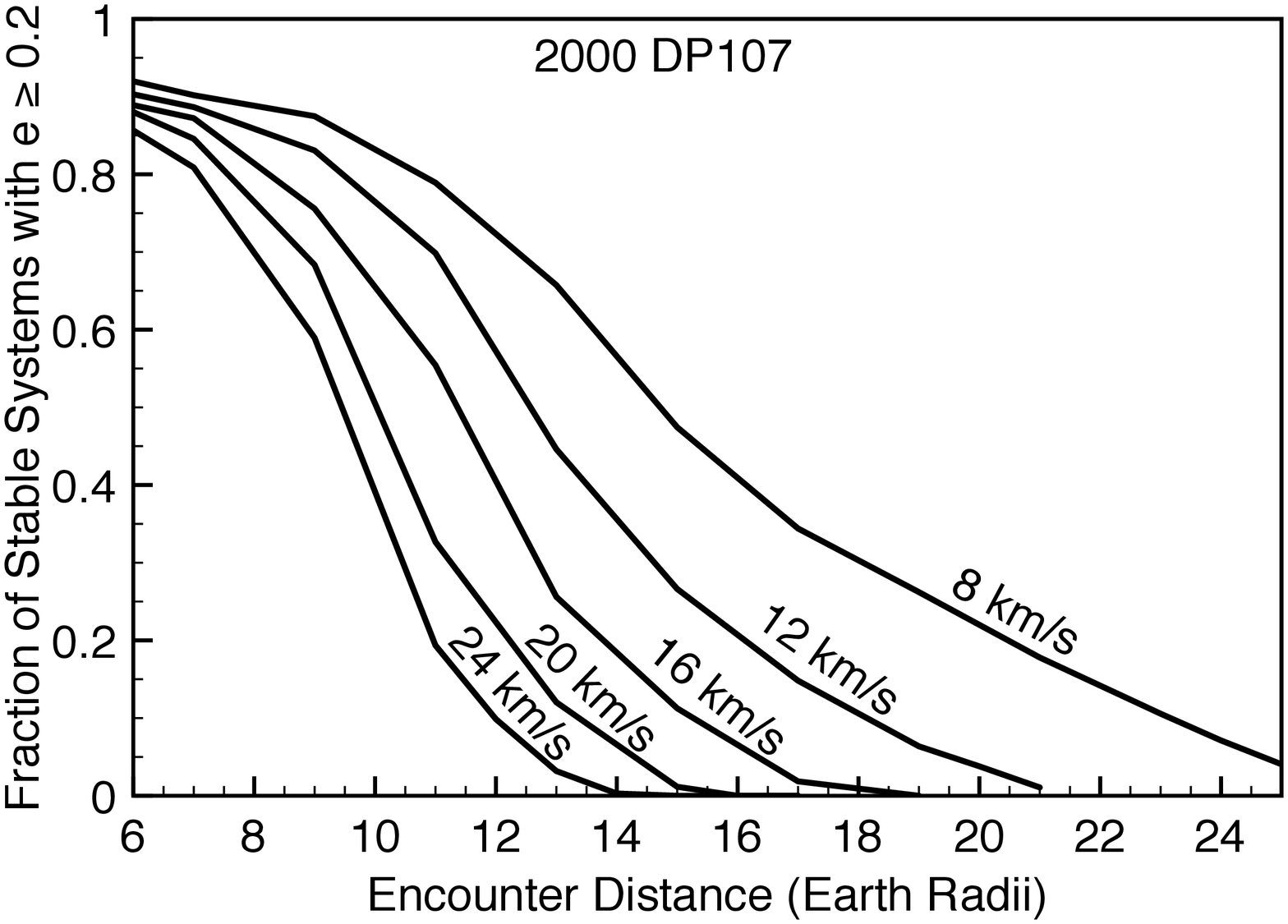}
	\includegraphics[scale=0.35]{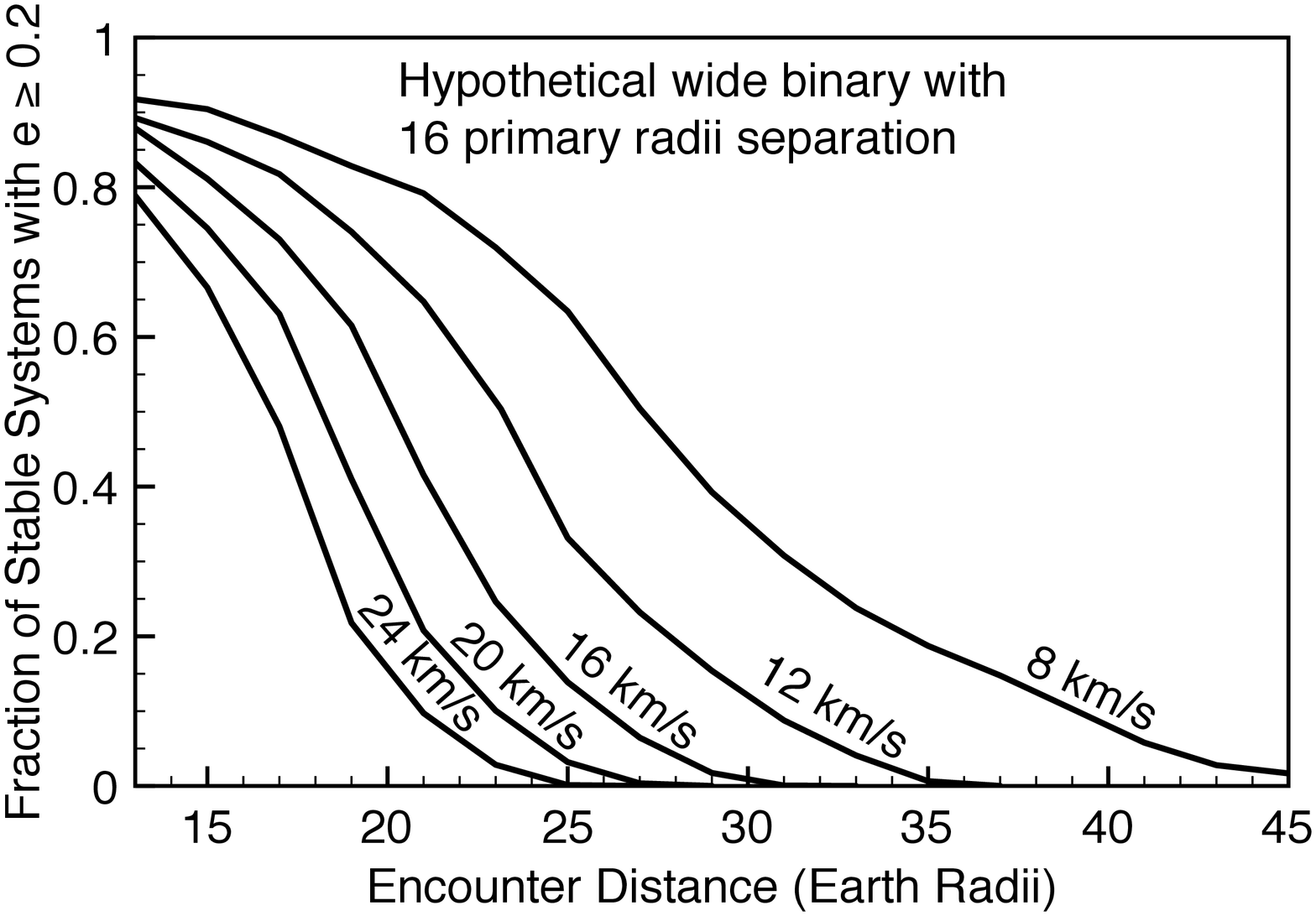}
	\caption{These plots show the characteristic behavior of close
	NEA binaries (i.e.\ 1999~KW4), moderately-separated NEA
	binaries (i.e.\ 2000~DP107), and wide NEA binaries 
        (represented by a hypothetical binary with a separation of 16 primary radii) 
		due to varying encounter velocities ($v_{\infty}$ =
	8$-$24 km~s$^{-1}$ with increments of 4 km~s$^{-1}$). The
	fraction of stable systems (no ejections nor collisions) with
	excited eccentricities ($e$ $\geq$ 0.2) is shown as a function
	of encounter distance. \label{vinfs}}
\end{figure}

The change in eccentricity is shown in Figure \ref{delta_e_plots},
with similar encounter distance and velocity ranges as shown in Figure
\ref{delta_a_plots}. Results from numerical simulations are given as
the mean value of the eccentricity increase of all stable encounters. 
Analytical and numerical results are in
agreement, especially for greater encounter velocities. For our test
NEA binaries (Table \ref{nea}) and using a typical $v_{\infty}$ of 12
km~s$^{-1}$, we also show the close encounter distances at which their
eccentricities can be excited to a value of at least 0.2 (Figure
\ref{neaexcite}) and how the distance at which eccentricity excitation
occurs is a function of binary separation (Figure \ref{a_excite}). The
eccentricity excitation distance is defined in Figure \ref{a_excite}
as the encounter distance where 50\% of stable encounters resulted in
binaries with $e$ $\geq$ 0.2. Since it is evident from this figure that there is a
relationship between the critical encounter distance $q$ and the
binary's semi-major axis $a$ (when expressed in units of primary
radii), we solve for its analytical relationship by rearranging
Equation \ref{delta_e}, and find
\begin{equation}
	\dfrac{q}{R_{\Earth}} \approx \dfrac{1}{R_{\Earth}} \left( \dfrac{1.89 M_{\Earth}}{\Delta e v_{\infty}} \right)^{1/2} \left( \dfrac{3G}{4 \pi \rho} \right)^{1/4} \left(\dfrac{a}{R_p}\right)^{3/4} \label{enc_eqn1}
\end{equation}
where $R_p$ is the primary's mass, $R_{\Earth}$ and $M_{\Earth}$ are the radius and mass
of Earth, $G$ is the gravitational constant, $\Delta e$ is the change
in eccentricity due to the encounter, $v_{\infty}$ is the encounter
velocity, and $\rho$ is the binary's density. 

As stated earlier, the
1.89 factor is due to averaging over all angles of encounter
geometries \citep{coll08}. This equation is valid for initially
circular binaries. The same $q \propto a^{3/4}$ relationship holds
for all binaries in our sample because for a uniform set of encounter
parameters (encounter velocity and perturber's radius and mass),
encounter strength (strong enough to excite $\Delta e = 0.2$), and
binary density (assuming NEA binaries have similar rubble pile
densities), the prefactor in front of the $a^{3/4}$ term is
constant. 
In convenient units,
\begin{equation}
	\frac{q}{R_\Earth} \approx 3.21 \left(\frac{0.2}{\Delta e} \times \frac{12\ {\rm km}~{\rm  s}^{-1}}{v_{\infty}}\right)^{1/2} \left(\frac{2\ {\rm g}~{\rm cm}^{-3}}{\rho}\right)^{1/4} \left(\frac{a}{R_p}\right)^{3/4} \label{enc_eqn2}
\end{equation}

We overplot this analytical relationship in Figure \ref{a_excite} and
find a good match to the numerical results. The agreement between
numerical and analytical estimates for 2003~YT1
is not as
good due to its higher density estimate (its nominal 
density is $\rho \sim$ 2.7 g cm$^{-3}$), 
since we assumed $\rho$ = 2 g cm$^{-3}$ in the calculation
of Equation \ref{enc_eqn2}. 

In Figure \ref{vinfs}, we show the
eccentricity excitation behavior due to $v_{\infty}$ = 8, 12, 16, 20,
and 24 km~s$^{-1}$ for specific observed NEA binaries: a close binary
(depicted by 1999~KW4), a moderately-separated binary (2000~DP107), and
a wide binary (represented by a hypothetical binary with a 16 primary
radii separation). As the value of $v_{\infty}$ decreases, a
greater fraction of stable systems have excited eccentricities due to
a planetary flyby at a given encounter distance. 
In addition, as the value of $v_{\infty}$ decreases, the final
eccentricities of binaries in our simulations increased. These effects
occur because a slower-passing perturber will have a longer encounter
duration with the binary and therefore cause a stronger perturbation
on the binary's mutual orbit.

\begin{figure}[htb]
	\centering
	\includegraphics[scale=0.35]{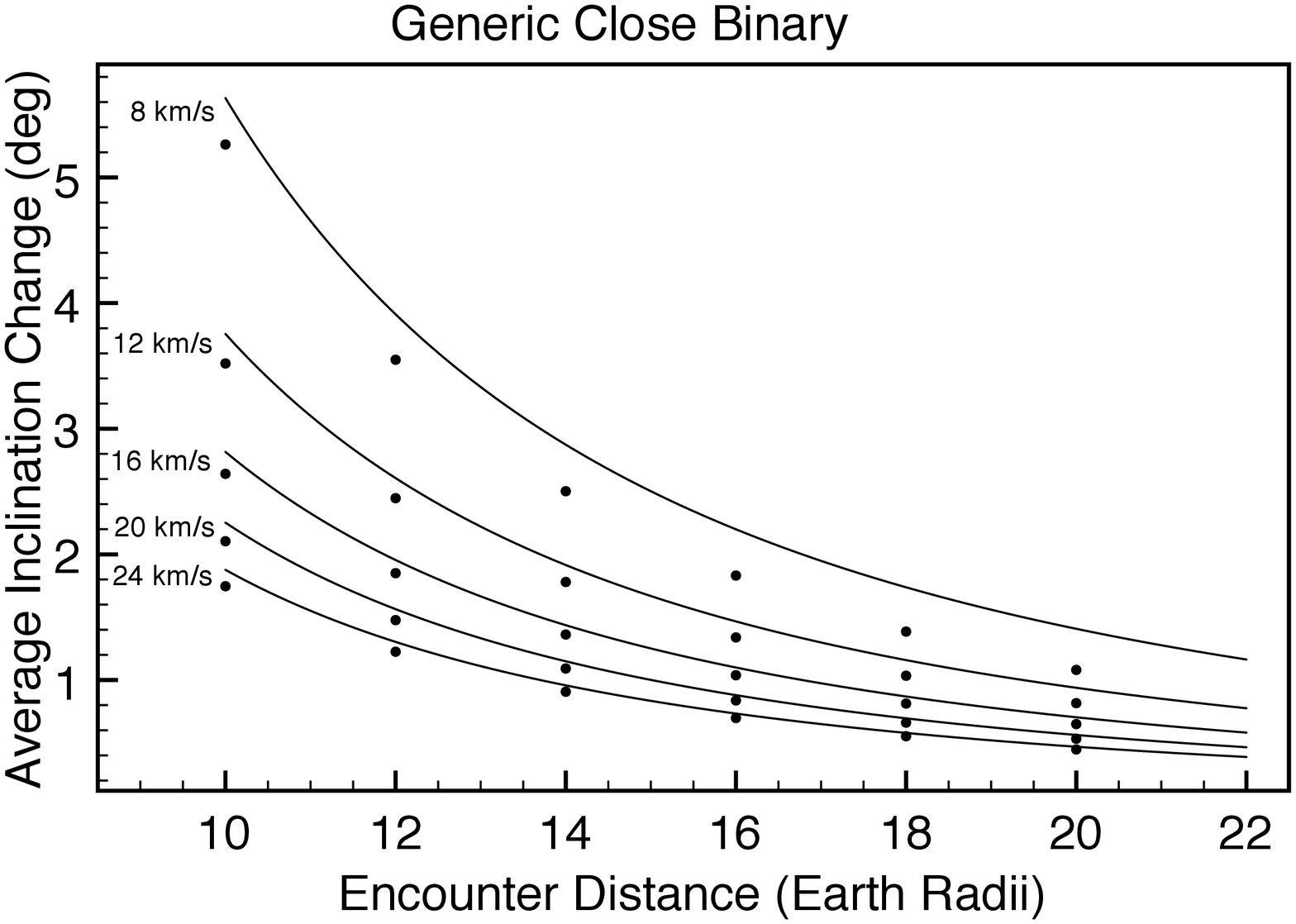}
	\includegraphics[scale=0.35]{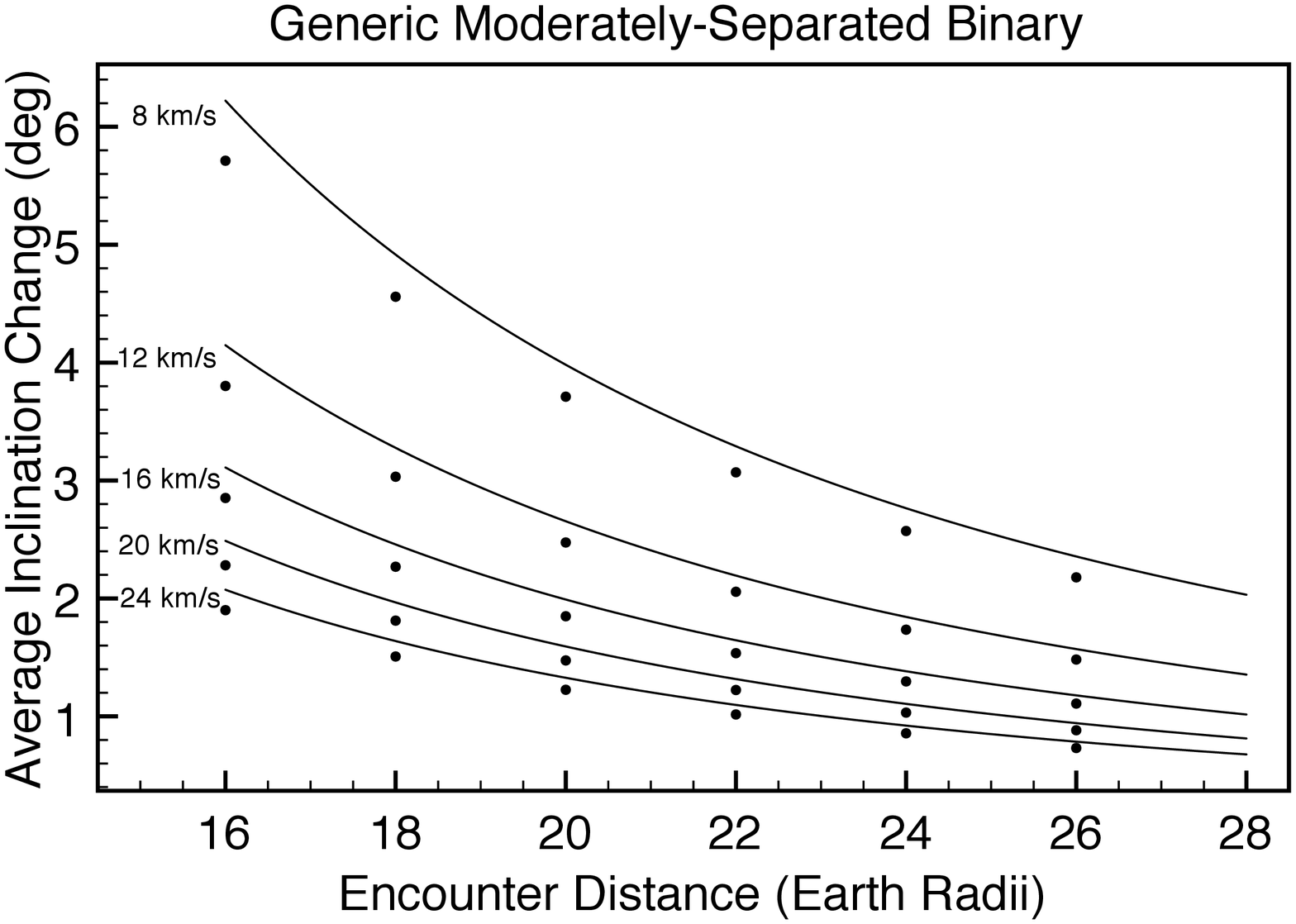}
	\includegraphics[scale=0.35]{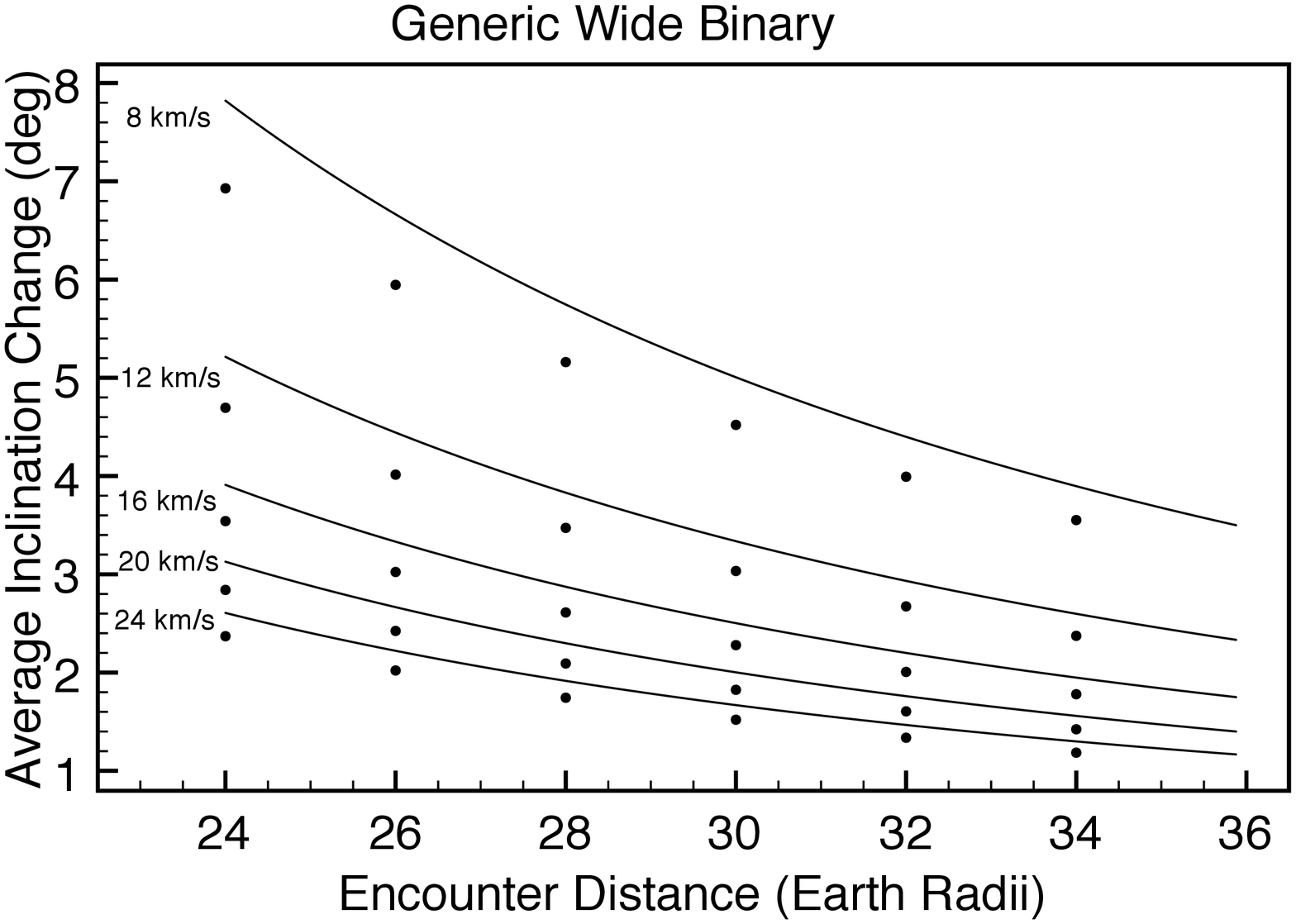}
	\caption{The change in a binary's inclination is shown as a
	function of encounter distance and velocity for 3 types of
	binaries: close binaries, moderately-separated binaries, and
	wide binaries. Results from numerical simulations are shown
	as dots and analytical calculations are depicted by
	solid lines. \label{delta_i_plots}}
\end{figure}


The change in a binary's orbital inclination is shown in Figure
\ref{delta_i_plots}, with similar encounter distances and velocities
as shown in Figures \ref{delta_a_plots} and
\ref{delta_e_plots}. Results from numerical simulations are given as
the mean value of the inclination change of all stable encounters.
 Comparison between output from numerical
simulations and analytical estimates provide a decent match. 
Possible reasons for the less-than-exact agreement between numerical
and analytical results include our choice of using the mean value to
represent the average change in inclination of stable encounters from
simulations as well as the practical limit on the number of encounter geometries
we were able to perform in our simulations. The analytical expressions
are also only valid in the limit of impulsive encounters, suggesting that
analytical results should provide better matches to simulations with
faster encounter velocities, which is consistent with our results.

These results have implications for binaries observed with excited
dynamical states. For instance, previous studies of 1999~KW4 
(heliocentric $a$ is 0.64 AU, $e$ is 0.69, $i$ is 38.89 degrees; 
current solar close approach distance is 0.2 AU) suggest
that the binary's close approaches to the Sun can excite its
rotational and orbital states \citep{ostr06,sche06}. Their numerical
simulations show that the binary's mutual orbit pole can be modified
by more than 0.5 degrees per pericenter passage for a pericenter
distance of 0.2 AU and by more than 1 degree for a pericenter distance
of 0.12 AU. 
From our simulations, we find that planetary encounters with an
Earth-mass body can perturb the binary orbit pole by a similar amount.
Therefore, even NEAs that do not come as close to the Sun as 1999~KW4
may exhibit excited rotational and orbital dynamics. For a generic
binary with a semi-major axis of 4 primary radii (same as 1999~KW4's
separation), our simulations using $v_{\infty}$ from 8$-$24 km
s$^{-1}$ show that a 0.5 degree shift in inclination can occur due to
a planetary encounter at distances of 18$-$28 $R_{\Earth}$ and a 1
degree shift can occur at 13$-$20 $R_{\Earth}$.

We briefly discuss instability trends seen in simulations with a
typical $v_{\infty}$ of 12 km~s$^{-1}$ for observed NEA binaries
(Table \ref{nea}). For nearly all encounter distances sampled in
simulations, ejections dominated over collisions in unstable
encounters. If planetary encounters close enough to disrupt a binary
have occurred, an ejected secondary is more likely to occur than a
collision between the primary and secondary.  
Therefore one would expect that planetary encounters can lead to the
formation of asteroid pairs in the near-Earth population, in addition
to the formation mechanisms responsible for pair production in the
main belt. The different routes towards the creation of asteroid pairs
are discussed in a companion paper by \citet{fang11ii}. 
Close binaries such as Didymos, 2002~CE26, 1999~KW4, and
2004~DC have similarly small component separations (2$-$5 $R_p$)
relative to their primary's radius $R_p$, and exhibit comparable
disruption statistics. Instabilities, including both ejections and
collisions, occurred at encounter distances of $\sim$5$-$7
$R_{\Earth}$ and less.  More moderately-spaced binaries such as
2000~DP107, 2003~YT1, and 1991~VH have comparable component
separations (5$-$9 $R_p$). For these binaries, simulation results
showed disruption occurring at encounter distances of $\sim$7$-$9
$R_{\Earth}$ and less.  Binaries with wide separations will have
mutual orbits that are more easily perturbed than tighter systems.

Given the significant orbital perturbations in binary systems that can be caused 
by planetary flybys, the next section examines the frequency of such encounters.

\section{Frequency of Planetary Encounters} \label{longterm}

\def\arraystretch{1.4}
\begin{deluxetable*}{lrrrrrrrrrrrrrrr}
\tablecolumns{16}
\tablecaption{Encounter Timescales (Myr) with Earth \label{times}}
\startdata
\hline \hline
\bf Name & \multicolumn{10}{l}{\bf Impact Parameter:} \\
	& 2$R_{\Earth}$	& 4$R_{\Earth}$ & 6$R_{\Earth}$ & 8$R_{\Earth}$ & 10$R_{\Earth}$ & 12$R_{\Earth}$ & 14$R_{\Earth}$ & 16$R_{\Earth}$ & 18$R_{\Earth}$ & 20$R_{\Earth}$ & 22$R_{\Earth}$ & 24$R_{\Earth}$ 
	& 26$R_{\Earth}$ & 28$R_{\Earth}$ & 30$R_{\Earth}$\\
\hline
2000~DP107 & 60.73  & 15.18 & 6.75  & 3.80  & 2.43  & 1.69  & 1.24 &  0.95 &  0.75 &  0.61 &  0.50 &  0.42 &  0.36 &  0.31 &  0.27 \\
1999~KW4 & 50.72  & 12.68 & 5.64  & 3.17  & 2.03  & 1.41  & 1.04 &  0.79 &  0.63 &  0.51 &  0.42 &  0.35 &  0.30 &  0.26 &  0.23 \\
2002~CE26 & 391.40 & 97.85 & 43.49 & 24.46 & 15.66 & 10.87 & 7.99 &  6.12 &  4.83 &  3.91 &  3.23 &  2.72 &  2.32 &  2.00 &  1.74 \\
2004 DC & 109.90 & 27.47 & 12.21 & 6.87  & 4.40  & 3.05  & 2.24 &  1.72 &  1.36 &  1.10 &  0.91 &  0.76 &  0.65 &  0.56 &  0.49 \\
2003~YT1 & 131.34 & 32.84 & 14.59 & 8.21  & 5.25  & 3.65  & 2.68 &  2.05 &  1.62 &  1.31 &  1.09 &  0.91 &  0.78 &  0.67 &  0.58 \\
Didymos & 46.30  & 11.58 & 5.14  & 2.89  & 1.85  & 1.29  & 0.94 &  0.72 &  0.57 &  0.46 &  0.38 &  0.32 &  0.27 &  0.24 &  0.21 \\
1991~VH & 72.86  & 18.21 & 8.10  & 4.55  & 2.91  & 2.02  & 1.49 &  1.14 &  0.90 &  0.73 &  0.60 &  0.51 &  0.43 &  0.37 &  0.32
\enddata
\tablenotetext{}{For each NEA binary, we show the encounter timescale
(in millions of years) with Earth for various impact parameters.}
\end{deluxetable*}

\subsection{Methods}

After investigating the orbital effects of a single planetary flyby on
a binary in Section \ref{nea_single}, we next determine the frequency
of such close planetary encounters with Earth for well-characterized
NEA binaries (Table \ref{nea}). The timescale of close encounters is
very dependent on the specific NEA's past evolutionary path to its
current heliocentric orbit (see Table \ref{nea} for current
heliocentric parameters), and different NEAs can have very different
close encounter histories. We do not backward-integrate the current
heliocentric orbits because the orbits become chaotic over time and it
is difficult to reconstruct their past history, even
statistically. Consequently, we follow the methods of \citet{bott02}
by integrating test particles from the three strongest source regions
($\nu_6$ secular resonance with Saturn, 3:1 mean-motion resonance with
Jupiter, and Mars-crossing regions) in or adjacent to the main belt
and by tracking the test particles as they migrate into near-Earth
space. These source regions are the main dynamical pathways to NEA
orbits. Integrations are performed using a hybrid
symplectic/Bulirsch-Stoer algorithm from {\em Mercury} \citep{cham99},
and include the Sun and all 8 planets in addition to 3000 test
particles per source (9000 test particles total). In the simulations, we ignore the effect of 
Yarkovsky. The Yarkovsky effect is an important perturbation
that moves asteroids into main source regions; however, once asteroids 
have reached the source regions, the effect of Yarkovsky is negligible
compared to the effect of strong resonances and planetary encounters.

After test particles have migrated into near-Earth space, we search
for particles whose orbital elements (semi-major axis $a$,
eccentricity $e$, and inclination $i$) closely match the current
orbital elements of the actual NEA binaries in our sample (Table
\ref{nea}). For each binary, we search for ten test particles 
closest in orbital element space to the binary's current heliocentric 
orbital elements using this scaling: 
$\delta a/a \sim \delta e \sim \delta \sin i$ \citep{vokr08}.
Binaries 2003 YT1 and 1999 KW4 have either high inclinations
and/or low semi-major axes, and these binaries proved to be difficult when
searching for matches with test particles originating from the 3
main source regions given by \citet{bott02}. 
In our sample of binaries, the median $\delta a$ is $\sim$0.014 AU. 
The matching test particles are further integrated until they become
dynamically unstable through ejection from the Solar System or collision
with a planet or the Sun.
For each NEA binary's matching test particles, we record the orbital
history with an Opik-type code similar to the procedure developed by
\citet{weth67} and implemented by \citet{fari92}, which allows us to
calculate the encounter velocity and intrinsic collisional probability
with Earth at each timestep.
We calculate the intrinsic collisional probability with Earth every
10,000 years from the time of injection in the source region up to the
most recent epoch at which the test particle's orbital elements matched
those of the actual NEA binary.  
As a result, we obtain an intrinsic collisional probability
value for each 10,000-year timestep in the history of each matching test particle.
Based on this ensemble of values we associate an average intrinsic
collisional probability to each test particle by averaging all non-zero values over time.
Finally, we associate an intrinsic collisional probability to each NEA binary
by averaging over all of its matching test particles.

The intrinsic collisional probability (in units of km$^{-2}$
yr$^{-1}$) can be multiplied by the square of an impact parameter to
obtain an encounter probability (in units of yr$^{-1}$), whose inverse
yields an approximate encounter timescale. These timescales represent
the average time between encounters with Earth since injection into
main belt source regions and their subsequent migration into
near-Earth space.

We also calculate the close encounter timescale using an analytical,
order-of-magnitude approach. \citet{chau95} give the approximate
timescale $t$
\begin{equation}
	t \sim \dfrac{2\tau_{\rm coll}R_{\Earth}^2}{b^2}
\label{nealife}
\end{equation}
between close encounters up to an impact parameter $b$ given an
asteroid's lifetime $\tau_{\rm coll}$ against collision with Earth of
radius $R_{\Earth}$. For a given asteroid size, \citet{stua04} provide
the total number of NEAs with at least that size as well as the
average interval between Earth impacts due to all asteroids of that
size taken collectively. Thus, with these numbers we can calculate
$\tau_{\rm coll}$ and the encounter timescale for an individual
asteroid.

\subsection{Results}

For each binary in our sample, Table \ref{times} gives encounter
timescales with Earth from numerical simulations for impact parameters ranging from 2$R_{\Earth}$
to 30$R_{\Earth}$. In this impact parameter range, encounter
timescales can range from $\sim$10$^5$ up to $\sim$10$^9$ years for the NEA binaries
considered here. Encounter probabilities can be obtained by taking the
inverse of these timescales, and then suitably scaled (the probability
varies as the square of the impact parameter) to a desired impact
parameter. Recall that due to gravitational focusing, the impact parameter $b$
is not the same as the closest encounter distance $q$, and are related
in this manner: $b^2 = q^2 (1 + (2 G M_{\Earth})/(q v_{\infty}^2))$,
where $G$ is the gravitational constant, $M_{\Earth}$ is the mass of
Earth, and $v_{\infty}$ is the encounter velocity at infinity.

Timescale estimates from the analytical approach agree with results
obtained from numerical integrations within an order of magnitude for all NEA
binaries in our sample. Differences in timescales can be
attributed to the fact that the analytical timescales do not take into
account a specific NEA's orbital trajectory as it migrates to
near-Earth space, which affects its history of planetary
encounters.

We repeat this analysis to determine encounter frequency with other
terrestrial planets, namely Mercury, Venus, and Mars. For a given
orbital effect such as increasing a binary's eccentricity to a given
value, we find that encounters with Venus are just as important as
encounters with Earth. For all binaries in our sample, the 
encounter probabilities with Venus and Earth are comparable; these encounter probabilities
for achieving a given change in the mutual orbit are a function of
encounter velocity, planet mass, and binary separation.

\section{Discussion and Conclusion} \label{conclusion}

Encounter timescales presented in Table \ref{times} show that some NEA
binaries in our sample (Table \ref{nea}) can typically encounter
Earth frequently enough at close-enough distances to
excite the binary's orbital elements, as shown by single-flyby
simulations in Section \ref{nea_single}. We have presented the
encounter timescales in this study, but the actual effects of
encounters are also dependent on typical encounter velocities, which
we plotted in Figure \ref{nea_vinf} for a generic NEA.  Consideration
of these factors as well as comparison to observed eccentricities are
discussed in a companion paper by \citet{fang11ii}.

We briefly address the possibility of repeat encounter passes. Repeat
passes, which are most relevant for binaries with short encounter
timescales and large binary separations, can increase or decrease the
binary's semi-major axis and eccentricity depending on its values
prior to the encounter. The strength of repeat passes depends on flyby
parameters such as encounter velocity. The net effect of repeat passes
is still an eccentric orbit, since it is rare for an eccentric binary
to undergo a planetary encounter and end up with near-zero
eccentricity.

Close planetary encounters have important implications for an NEA
binary's evolution, since flybys can disrupt main evolutionary
processes in a binary by expanding or contracting the mutual
orbit. For a typical NEA binary with a separation of $\sim$4 primary
radii, we find that the semi-major axis increases on average 60\% of
the time for encounter distances from 2$-$10 $R_{\Earth}$ and
encounter velocities from 8$-$24 km~s$^{-1}$. Since flybys can
increase or decrease the semi-major axis of the binary's orbit, tidal
evolution can strengthen (if the semi-major axis decreases) or weaken
(if the semi-major axis increases) since the tidal torque scales as
the binary separation to the sixth power. 
When the semi-major axis is modified, the mean motion of the secondary changes and an initially
spin-locked secondary may become asynchronous\footnote[1]{
In this paper, binaries with an absence of spin-orbit synchronism are
called {\em asynchronous binaries}.  Binaries with a secondary spin
period synchronized to the mutual orbit period are called {\em
synchronous binaries}.  Binaries with both primary and secondary spin
periods synchronized to the mutual orbit period are called {\em doubly
synchronous binaries}.  Most NEA binaries are {\em synchronous}. Note
that our terminology is different from that of \citet{prav07}, who
used the term ``asynchronous binaries" for binaries with spin-orbit
synchronization. If generalization to systems with more than one 
satellite is needed, we affix the terms {\em synchronous} and 
{\em asynchronous} to the satellites being considered. \\
}. Asynchronization can
also occur by a change in the secondary's rotation rate due to the
flyby. The loss of spin-lock would imply that radiative perturbations
dependent on synchronization such as binary YORP (BYORP) would be shut
down. Lastly, planetary encounters are another mechanism that may
form near-Earth asteroid pairs. The evolution
of binaries under various influences, including the planetary model
presented here, is further discussed in a companion paper by \citet{fang11ii}.

To summarize, in this study we have used numerical integrations and
analytical expressions to investigate the effects of planetary
encounters on NEA binaries. We found the encounter distances at which
flybys can increase the orbital semi-major axis, eccentricity, and
inclination for a variety of encounter velocities. There is reasonable
agreement between results obtained from simulations and analytical
methods. The possible outcomes, including collisions, ejections, and
stable encounters, are discussed for close binaries,
moderately-separated binaries, and wide binaries. We have also used
N-body integrations to examine the past evolutionary histories of NEAs
as they migrated from main belt source regions into near-Earth
space. From these simulations, we calculated encounter probabilities
and timescales for all NEA binaries in our sample for different impact
parameters. These encounter timescales provide rough agreement with
analytical estimates, which do not take into account individual NEAs'
past orbital histories. Lastly, planetary encounters have important
implications for evolutionary processes such as tidal and BYORP
mechanisms.

\acknowledgments

We thank Bill Bottke and Ben Collins for useful discussions. We are
also grateful to the reviewer for helpful comments. This work
was partially supported by NASA Planetary Astronomy grant NNX09AQ68G.

\bibliographystyle{apj}
\bibliography{binaries}

\begin{thebibliography}{34}
\expandafter\ifx\csname natexlab\endcsname\relax\def\natexlab#1{#1}\fi

\bibitem[{{Asphaug} \& {Benz}(1996)}]{asph96}
{Asphaug}, E. \& {Benz}, W. 1996, \icarus, 121, 225

\bibitem[{{Benner} {et~al.}(2010){Benner}, {Margot}, {Nolan}, {Giorgini},
  {Brozovic}, {Scheeres}, {Magri}, \& {Ostro}}]{benn10}
{Benner}, L.~A.~M., {Margot}, J.~L., {Nolan}, M.~C., {Giorgini}, J.~D.,
  {Brozovic}, M., {Scheeres}, D.~J., {Magri}, C., \& {Ostro}, S.~J. 2010, in
  Bulletin of the American Astronomical Society, Vol.~42, AAS/Division for
  Planetary Sciences Meeting Abstracts \#42, 1056--+

\bibitem[{{Binzel} {et~al.}(2010){Binzel}, {Morbidelli}, {Merouane}, {DeMeo},
  {Birlan}, {Vernazza}, {Thomas}, {Rivkin}, {Bus}, \& {Tokunaga}}]{binz10}
{Binzel}, R.~P., {Morbidelli}, A., {Merouane}, S., {DeMeo}, F.~E., {Birlan},
  M., {Vernazza}, P., {Thomas}, C.~A., {Rivkin}, A.~S., {Bus}, S.~J., \&
  {Tokunaga}, A.~T. 2010, \nat, 463, 331

\bibitem[{{Bottke} \& {Melosh}(1996{\natexlab{a}})}]{bott96b}
{Bottke}, W.~F. \& {Melosh}, H.~J. 1996{\natexlab{a}}, \nat, 381, 51

\bibitem[{{Bottke} {et~al.}(2002){Bottke}, {Morbidelli}, {Jedicke}, {Petit},
  {Levison}, {Michel}, \& {Metcalfe}}]{bott02}
{Bottke}, W.~F., {Morbidelli}, A., {Jedicke}, R., {Petit}, J.-M., {Levison},
  H.~F., {Michel}, P., \& {Metcalfe}, T.~S. 2002, \icarus, 156, 399

\bibitem[{{Bottke} \& {Melosh}(1996{\natexlab{b}})}]{bott96a}
{Bottke}, Jr., W.~F. \& {Melosh}, H.~J. 1996{\natexlab{b}}, \icarus, 124, 372

\bibitem[{{Chambers}(1999)}]{cham99}
{Chambers}, J.~E. 1999, \mnras, 304, 793

\bibitem[{{Chauvineau} \& {Farinella}(1995)}]{chau95}
{Chauvineau}, B. \& {Farinella}, P. 1995, \icarus, 115, 36

\bibitem[{{Collins} \& {Sari}(2008)}]{coll08}
{Collins}, B.~F. \& {Sari}, R. 2008, \aj, 136, 2552

\bibitem[{{{\'C}uk}(2007)}]{cuk07}
{{\'C}uk}, M. 2007, \apjl, 659, L57

\bibitem[{{{\'C}uk} \& {Burns}(2005)}]{cuk05}
{{\'C}uk}, M. \& {Burns}, J.~A. 2005, \icarus, 176, 418

\bibitem[{{Fang} {et~al.}(2011){Fang}, {Margot}, {Brozovic}, {Nolan}, {Benner},
  \& {Taylor}}]{fang11}
{Fang}, J., {Margot}, J., {Brozovic}, M., {Nolan}, M.~C., {Benner}, L.~A.~M.,
  \& {Taylor}, P.~A. 2011, \aj, 141, 154

\bibitem[{{Fang} \& {Margot}(2011)}]{fang11ii}
{Fang}, J. \& {Margot}, J.~L. 2011, \aj

\bibitem[{{Farinella}(1992)}]{fari92b}
{Farinella}, P. 1992, \icarus, 96, 284

\bibitem[{{Farinella} \& {Chauvineau}(1993)}]{fari93}
{Farinella}, P. \& {Chauvineau}, B. 1993, \aap, 279, 251

\bibitem[{{Farinella} \& {Davis}(1992)}]{fari92}
{Farinella}, P. \& {Davis}, D.~R. 1992, \icarus, 97, 111

\bibitem[{{Heggie} \& {Rasio}(1996)}]{hegg96}
{Heggie}, D.~C. \& {Rasio}, F.~A. 1996, \mnras, 282, 1064

\bibitem[{{Margot} {et~al.}(2008){Margot}, {Taylor}, {Nolan}, {Howell},
  {Ostro}, {Benner}, {Giorgini}, {Magri}, \& {Carter}}]{marg08}
{Margot}, J., {Taylor}, P.~A., {Nolan}, M.~C., {Howell}, E.~S., {Ostro}, S.~J.,
  {Benner}, L.~A.~M., {Giorgini}, J.~D., {Magri}, C., \& {Carter}, L.~M. 2008,
  in Bulletin of the American Astronomical Society, Vol.~40, AAS/Division for
  Planetary Sciences Meeting Abstracts \#40, 433--+

\bibitem[{{Margot} {et~al.}(2002){Margot}, {Nolan}, {Benner}, {Ostro},
  {Jurgens}, {Giorgini}, {Slade}, \& {Campbell}}]{marg02}
{Margot}, J.~L., {Nolan}, M.~C., {Benner}, L.~A.~M., {Ostro}, S.~J., {Jurgens},
  R.~F., {Giorgini}, J.~D., {Slade}, M.~A., \& {Campbell}, D.~B. 2002, Science,
  296, 1445

\bibitem[{{Nesvorn{\'y}} {et~al.}(2010){Nesvorn{\'y}}, {Bottke},
  {Vokrouhlick{\'y}}, {Chapman}, \& {Rafkin}}]{nesv10}
{Nesvorn{\'y}}, D., {Bottke}, W.~F., {Vokrouhlick{\'y}}, D., {Chapman}, C.~R.,
  \& {Rafkin}, S. 2010, \icarus, 209, 510

\bibitem[{{Nolan} {et~al.}(2004){Nolan}, {Howell}, \& {Miranda}}]{nola04}
{Nolan}, M.~C., {Howell}, E.~S., \& {Miranda}, G. 2004, in Bulletin of the
  American Astronomical Society, Vol.~36, AAS/Division for Planetary Sciences
  Meeting Abstracts \#36, 1132--+

\bibitem[{{Ostro} {et~al.}(2006)}]{ostr06}
{Ostro}, S.~J. {et~al.} 2006, Science, 314, 1276

\bibitem[{{Pravec} \& {Harris}(2007)}]{prav07}
{Pravec}, P. \& {Harris}, A.~W. 2007, Icarus, 190, 250

\bibitem[{{Pravec} {et~al.}(2006)}]{prav06}
{Pravec}, P. {et~al.} 2006, \icarus, 181, 63

\bibitem[{{Richardson} {et~al.}(1998){Richardson}, {Bottke}, \&
  {Love}}]{rich98}
{Richardson}, D.~C., {Bottke}, W.~F., \& {Love}, S.~G. 1998, \icarus, 134, 47

\bibitem[{{Scheeres} {et~al.}(2006){Scheeres}, {Fahnestock}, {Ostro}, {Margot},
  {Benner}, {Broschart}, {Bellerose}, {Giorgini}, {Nolan}, {Magri}, {Pravec},
  {Scheirich}, {Rose}, {Jurgens}, {De Jong}, \& {Suzuki}}]{sche06}
{Scheeres}, D.~J., {Fahnestock}, E.~G., {Ostro}, S.~J., {Margot}, J.~L.,
  {Benner}, L.~A.~M., {Broschart}, S.~B., {Bellerose}, J., {Giorgini}, J.~D.,
  {Nolan}, M.~C., {Magri}, C., {Pravec}, P., {Scheirich}, P., {Rose}, R.,
  {Jurgens}, R.~F., {De Jong}, E.~M., \& {Suzuki}, S. 2006, Science, 314, 1280

\bibitem[{{Scheeres} {et~al.}(2004){Scheeres}, {Marzari}, \& {Rossi}}]{sche04}
{Scheeres}, D.~J., {Marzari}, F., \& {Rossi}, A. 2004, \icarus, 170, 312

\bibitem[{{Sharma} {et~al.}(2006){Sharma}, {Jenkins}, \& {Burns}}]{shar06}
{Sharma}, I., {Jenkins}, J.~T., \& {Burns}, J.~A. 2006, \icarus, 183, 312

\bibitem[{{Shepard} {et~al.}(2006)}]{shep06}
{Shepard}, M.~K. {et~al.} 2006, \icarus, 184, 198

\bibitem[{{Stuart} \& {Binzel}(2004)}]{stua04}
{Stuart}, J.~S. \& {Binzel}, R.~P. 2004, \icarus, 170, 295

\bibitem[{{Taylor} {et~al.}(2008){Taylor}, {Margot}, {Nolan}, {Benner},
  {Ostro}, {Giorgini}, \& {Magri}}]{tayl08b}
{Taylor}, P.~A., {Margot}, J.~L., {Nolan}, M.~C., {Benner}, L.~A.~M., {Ostro},
  S.~J., {Giorgini}, J.~D., \& {Magri}, C. 2008, LPI Contributions, 1405, 8322

\bibitem[{{Vokrouhlick{\'y}} \& {Nesvorn{\'y}}(2008)}]{vokr08}
{Vokrouhlick{\'y}}, D. \& {Nesvorn{\'y}}, D. 2008, \aj, 136, 280

\bibitem[{{Walsh} \& {Richardson}(2006)}]{wals06}
{Walsh}, K.~J. \& {Richardson}, D.~C. 2006, \icarus, 180, 201

\bibitem[{{Wetherill}(1967)}]{weth67}
{Wetherill}, G.~W. 1967, \jgr, 72, 2429

\end{thebibliography}

\end{document}